\documentclass[aps,prl,twocolumn,showpacs,superscriptaddress,groupedaddress]{revtex4}
\usepackage{amsmath}
\usepackage{graphicx}
\usepackage{natbib}
\usepackage[usenames]{color}
\usepackage[T1]{fontenc}
\usepackage{textcomp}

\begin{document}


\title{Scarcity may promote cooperation in populations of simple agents}

\author{R.~J.~Requejo}
\author{J.~Camacho}
\affiliation{Departament de F\'isica, Universitat Aut\`onoma de Barcelona, Campus UAB, E-08193 Bellaterra, Spain.}
\email {juan.camacho@uab.es}

\date{\today}

\begin{abstract}
In the study of the evolution of cooperation, resource limitations are usually assumed just to provide a finite population size. Recently, however, it has been pointed out that resource limitation may also generate dynamical payoffs able to modify the original structure of the games. Here we study analytically a phase transition from a homogeneous population of defectors when resources are abundant, to the survival of unconditional cooperators when resources reduce below a threshold. To this end, we introduce a model of simple agents, with no memory or the ability of recognition, interacting in well-mixed populations. The result might shed light on the role played by resource constraints on the origin of multicellularity.
\end{abstract}


\maketitle

Cooperation is common in nature in all levels of biological organization \cite{hammerstein:2003}, and it is considered to have played a key role in the evolutionary appearance of higher selective units, such as eukaryotic cells or multicellular life, from simpler components \cite{maynard-smith:1995a}. However, its widespread abundance is intriguing because cooperators are vulnerable to exploitation by defectors \cite{darwin:1859}, as detected early on \cite{hamilton:1964a,hamilton:1964b}. Since then, several mechanisms have been found allowing cooperative behaviors to survive such as network structure, group selection, direct and indirect reciprocity or tag-based donation \cite{riolo:2001,nowak:2006b}. Behavioral mechanisms --the latter three are examples of such mechanisms-- require players to have some ability to avoid the exploitation from defectors, such as memory or the capacity to recognise the co-player \cite{riolo:2001,nowak:2006b}. As a result, simple agents without these abilities, such as unconditional cooperators, are not expected to survive in well-mixed populations. 


Aside from a few examples \cite{wakano:2007,dobramysl:2008,hauert:2006a,melbinger:2010}, the role played by the limitation of resources in most evolutionary game theoretical studies on the origin and persistence of cooperation has been just to impose a constant population size \cite{trivers:1971,axelrod:1981,nowak:1998,nowak:2006b,riolo:2001,traulsen:2003,gomez-gardenes:2007,roca:2009}. Recently, however, we have put forward a new viewpoint where the interacting players are set into a nonequilibrium context \cite{requejo:2011,requejo:2012,requejo:2012b}. The environment is considered explicitly by introducing a resource flux into the system that drives it away from equilibrium. This standpoint leads to unexpected outcomes, such as that resource limitation allows for stable coexistence between unconditional cooperators and defectors, and even dominance of cooperation, in well-mixed populations playing an a priori Prisoner's Dilemma (PD) game. This happens due to a self-organizing process involving the environment which generates dynamical payoffs transforming the original PD structure into a neutral game. 


One of the main results of the analysis performed in ref.\cite{requejo:2012,requejo:2012b} is that a well-mixed population of unconditional cooperators extinguishes for infinite resources (where the system plays a PD game) but may survive for some parameter values when resources are finite (where the game is not a PD anymore). This suggests the possibility of a transition from a population of only defectors when resources are abundant to a population containing cooperators for more stringent environments. The existence of this transition should have great interest, since it would provide a resource-based mechanism preventing the spread of defectors and thus may shed light on the conditions under which cooperators could survive during the evolutionary process. Indeed, the survival of cooperative strains has been 
recently observed experimentally in yeast (\emph{S.Cerevisiae}) \cite{jansen:2005,wenger:2011,koschwanez:2011} and bacterial (\emph{E.Coli}) \cite{notley-mcrobb:1999,maharjan:2006} cultures, and has also been found in a model for the survival 
of aerobic cells inside anaerobic cultures \cite{pfeiffer:2001,pfeiffer:2003}. The models depicted in refs.\cite{requejo:2012,requejo:2012b}, however, do not yield such a transition: in these models we considered that the population was ruled by a resource limiting reproduction, and that deaths occurred at a constant rate, so that the limiting resource influx determined the population size; as it was thoroughly discussed, a reduction in the resource flux just decreased the size of the population in the same proportion, but it did not modify its composition. 

Our aim here is to devise a scenario where the selection pressure drived by resource limitation combined with the nonlinearities induced by this resource limitation in the interactions among players may lead to a transition of the type discussed above. In this scenario, we  assume a limiting resource that constraints reproduction in a population of constant size due, for instance to space constraints. The plaussibility of the latter assumption is discussed at the end of the paper.
The model developed here is a stylized one inspired in the model of ref. \cite{requejo:2012b}, which consists of an evolving
population of self-replicating individuals that receive resources from the environment and exchange resources during interactions. In order to avoid the effect of spatial structure and focus on the effect of resource constraints, we consider a well-mixed population.
No memory, learning abilities or any other sensory inputs are assumed. Each individual $i$ is represented by its amount of resources, $E_i$, which in this simplified model is either 0 or 1, and its strategy, namely cooperate (C) or defect (D). Its amount of resources may be interpreted as the amount that belongs to it independently of how (it may be in its surroundings, for instance).
Each defector attacks at a rate $\alpha$ per unit time to individuals chosen at random 
and steals its internal resources. To do so, the defector must have internal resources greater than 0 (i.e. $E_i=1$), otherwise it does not attack. In every interaction, the defector loses its unit of resources with probability $q$, which can thus be seen as the average cost paid by a defector in an interaction. 
If the interaction partner has no resources, no reward is obtained. Cooperators do nothing, they just eventually suffer from defectors' attacks. 
We assume that behaviors are inherited without mutation and represent physiologic or morphologic characteristics intrinsic to individuals which cannot be modified by choice. 

Each individual receives from the environment $\gamma$ units of resources per unit time independently of its strategy, thus not modifying the interaction payoff structure. When an individual with internal resources $E_i=1$ receives an extra unit of resources it splits into two identical copies, each one with $E_i=1$. Along with reproduction, we assume that players die with a probability $f$, independently of its strategy, in such a way that the number of individuals in the population remains constant. Therefore, resource allocation, reproduction and death rules are equal for both cooperators and defectors, being the strategy the only difference.

Let us note that, in this model, an increase in the environmental resource supply is represented by an increase in $\gamma$, the amount of per capita resources obtained by individuals. This contrasts with the model in refs. \cite{requejo:2012,requejo:2012b}, where an increase of resources leads to a proportional increase in the population size while keeping the same per capita value.

We consider simultaneous interactions and large populations so that we can make a continuum approach. We denote by $c_0$ and $c_1$ the fraction of cooperators with internal resources 0 and 1, and $d_1$ and $d_0=1-c_0-c_1-d_1$ the fraction of defectors with internal resources 1 and 0, respectively. The equations governing the evolution of cooperators are the following
\begin{eqnarray}
\label{c1}
\frac{dc_0}{dt}&=& \alpha c_1 d_1 - \gamma c_0 - f c_0 \\
\label{c2}
\frac{dc_1}{dt}&=& - \alpha c_1 d_1 +  \gamma(c_0 + c_1) - f c_1 
\end{eqnarray}
The $\alpha c_1 d_1$ term shows the fraction of cooperators $C_1$ that lose their internal resource unit after the attack of defectors (the latter pertaining to the population $d_1$); these individuals move from population $c_1$ to $c_0$. The term in $\gamma c_0$ quantifies the fraction of individuals $C_0$ that change to population $c_1$ after getting a unit of resources from the environment. In addition, individuals in population $c_1$ that receive resources from the environment replicate, thus increasing the $c_1$ population. The terms $f c_i$ describe the fraction of individuals dying in each population per unit time. 

To describe the evolution of defectors is enough to write the equation for population $d_1$ because $d_0$ is just the remaining fraction of the whole population. The dynamic equation for $d_1$ is 
\begin{equation}
\label{d1}
\frac{d d_1}{dt}= -\alpha q d_1 + \alpha c_1 d_1 + \gamma(d_0+d_1) - f d_1 \\.
\end{equation}
The terms related to deaths and resource allocation from the environment are analogous as for cooperators. The interaction term is as follows. On the one hand, with probability $q$ individuals $D_1$ lose its resource unit when interacting with individuals $C_0, D_0$ and $D_1$; this leads to a decrease in the population of $d_1$ in an amount $\alpha q d_1(c_0+d_0+d_1)$. On the other hand, when interacting with individuals $C_1$, individuals $D_1$ sequester their resource unit; therefore, either the population of $D_1$ does not change, with probability $q$, or it increases due to reproduction at a rate $\alpha c_1 d_1 (1-q)$. 

To complete the equations of the model, we need an expression for the death rate $f$. In order to have a constant population size, the frequency of deaths must equal the frequency of reproductions. This leads to
\begin{equation}
\label{f}
f= \gamma(c_1+d_1) + \alpha (1-q)(c_1+d_1)d_1 \\.
\end{equation}
The first term denotes reproductions due to resource allocation and the second one to reproduction of $D_1$ individuals when attacking individuals with $E_i=1$ and not paying the cost. Eqs. (\ref{c1})--(\ref{f}) are the equations of our model.
They can be further simplified by noticing that one can divide all the equations by parameter $\gamma$ and absorb it into the time parameter; therefore, there are just two dimensionless parameters in the model, $q$ and $\beta=\alpha/\gamma$. 
A large $\beta$ value indicates either large defector attack rates or small resource influxes from the environment; conversely, large resource influxes or small attacking rates yield small $\beta$ values. The dimensionless equations are the same Eqs (\ref{c1})--(\ref{f}) replacing $\alpha$ by $\beta$, and $\gamma$ by 1.

\begin{figure}
	\centering
   \includegraphics[width=84mm]{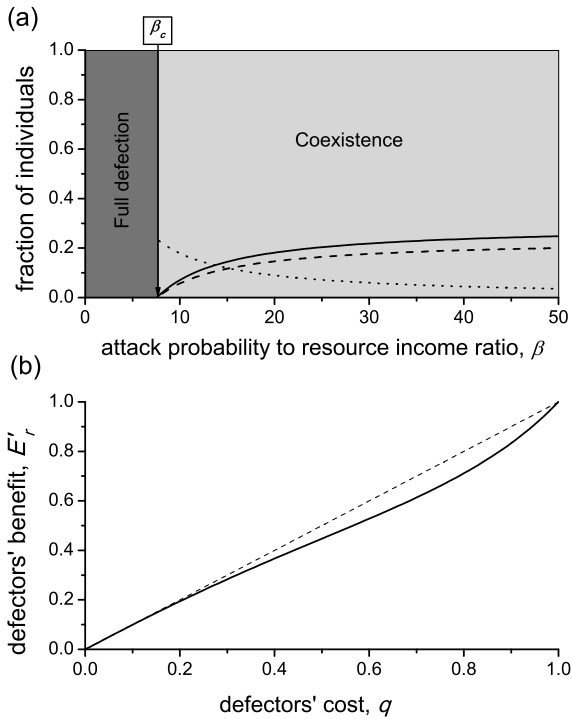}
\caption{\label{fig1} 
(a) Phase transition for q=0.5.  The fraction of cooperators $c_0$, $c_1$ and defectors $d_1$ (see text) above the threshold are denoted with solid, dashed and dotted lines, respectively. Below the critical value $\beta_c=7.58$ cooperators die out. (b) Defectors' benefit versus costs in coexistence states. It equals the frequency rate function $f(q)$ (see text). The dashed line $E'_r=q$ is a guide to the eye.}
\end{figure}

The numerical resolution of the model shows that the system is attracted to a globally stable fixed point independent of initial conditions. Depending on the parameter values, the final fate is either a population of defectors (an expected solution) or, interestingly, a stable mixture of cooperators and defectors. Remarkably enough, for fixed $q$, small $\beta$ values, i.e. large resource influxes, provide a population of just defectors, but when $\beta$ exceeds a critical value $\beta_c$ a mixed state appears, thus providing a smooth phase transition from defective states to mixed states as resources become scarce (see Fig.~\ref{fig1}a). The existence of stable mixed states in the model may be explained in terms of the overexploitation mechanism discussed in ref.~\cite{requejo:2012}: an excess of defectors may reduce cooperators' resource contents and, as a result, the average reward obtained by defectors; eventually, rewards decrease below costs and cooperators recover. Interestingly, we can obtain simple analytical expressions for the composition of the mixed state as a function of parameter $\beta$ above the threshold: 
\begin{equation}
\label{eq:mixed}
c_i= a_i (1-\frac{\beta_c}{\beta}),  \hspace{1cm} d_1=\frac{a_2}{\beta},
\end{equation}
with $a_i$ and $\beta_c$ functions of parameter $q$. 

Remarkably, the dynamics in coexistence states selforganizes defectors' rewards to be (almost) equal to costs thus turning the payoff matrix to neutral. According to the model, the payoff matrix for an average interaction is
\begin{center}
\begin{equation}
\label{(2.2)}
\bordermatrix{\text{}&$C$&$D$&\cr
                $C$&0 & - E'_r\cr
                $D$& E'_r-E_c & - E_c\cr}
\end{equation}
\end{center}
with $E'_r$ the average reward obtained when a defector attacks a cooperator, and $E_c=q$ the average cost paid when a defector attacks. Then, the average reward received by defectors when interacting with cooperators is $E'_r=c_1/(c_0+c_1)$. Eq.~(\ref{eq:mixed}) shows that $E'_r= a_1/(a_0+a_1)$ and then it is a function dependent only on $q$, and not on $\beta$. Fig.~\ref{fig1}b displays the reward $E'_r$ as obtained numerically versus the cost $q$ showing that $E'_r\simeq q$. They are not exactly equal because, as explained in Eq.~(4) of ref.~\cite{requejo:2012}, they may differ when death frequencies $f$ are not small compared with resource intake. In this model, $f$ cannot be arbitrarily chosen because of the constant population condition. Indeed, Eqs.~\ref{c1}--\ref{c2} readily show that $f=c_1/(c_0+c_1)=E'_r$ and then Fig.~\ref{fig1}b also displays $f(q)$. One observes that $f$ is generally of order 1 (this is the cause of the small deviations found in Fig.~\ref{fig1}b). At small $q$, however, $f$ is also small and $E'_r$ and $q$ match perfectly. 

\begin{figure}
\includegraphics*[width=8cm]{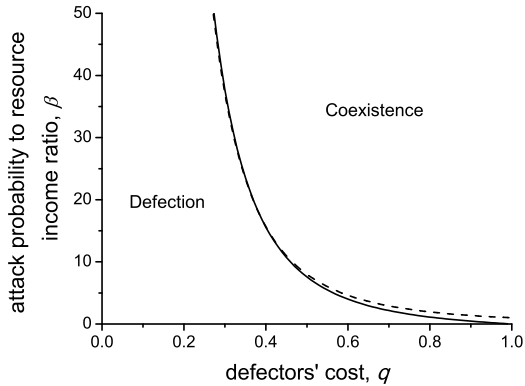}
\caption{\label{fig2} 
Phase diagram. The solid line indicates the numerical solution, the dashed line the analytical approximation $\beta_c=q^{-3}$. Cooperation is favoured at large $\beta$, i.e. small resource fluxes or large attacking rates (see text). }
\end{figure}

One can further study the transition by drawing a phase diagram 
$\beta-q$ with the regions where each behavior dominates. It is possible to obtain an analytical expression for the critical curve $\beta_c(q)$ by performing a stability analysis. To do so, let us recall that for a fixed point to be stable in three dimensions the trace and determinant of the Jacobian matrix must be negative.
Our model system (\ref{c1})--(\ref{f}) has at least two fixed points, corresponding to pure populations of cooperators and defectors: (A) $c_1=1$ (the remaining variables equal to 0), and (B) $c_0=c_1=0$ and $d_1\neq 0$ obeying, according to Eq.~\ref{d1},
\begin{equation}
\label{d1bis}
d_1(f+\beta q)=1. 
\end{equation}
As we know, it may also have a mixed fixed point, given by Eq.~\ref{eq:mixed}, but it need not be considered for our present purpose. Linear stability of fixed point A leads to the jacobian matrix
\begin{equation}
\begin{pmatrix}
\displaystyle -2  &
\displaystyle 0 &
\displaystyle  \beta \\[4mm]
\displaystyle 1  &
\displaystyle -1 &
\displaystyle  -1-2\beta+\beta q \\[4mm]
\displaystyle -1  &
\displaystyle -1 &
\displaystyle  \beta (1-q)-1
\end{pmatrix},
\end{equation}
with determinant $D=4\beta (1-q)>0$. The positive sign shows that at least one of its three eigenvalues is positive. Then point A is always unstable for $q<1$ and cooperators never occupy the whole population. Fixed point B provides the following jacobian matrix
\begin{equation}
\begin{pmatrix}
\displaystyle -1-f_*  &
\displaystyle \beta d_1^* &
\displaystyle 0 \\[4mm]
\displaystyle 1  &
\displaystyle 1-\beta d_1^*-f^* &
\displaystyle  0 \\[4mm]
\displaystyle -1  &
\displaystyle -1-d_1^*\frac{df}{dc_1}+\beta d_1^*&
\displaystyle  -\beta q-f_*-d_1^*\frac{df}{dd_1}
\end{pmatrix},
\end{equation}
where $d_1^*$ and $f_*$ are the values of these quantities in fixed point B. 
To be compact, let us call $J_{33}=-(\beta q+f_*+d_1^*(1+2\beta(1-q)d_1^*))<0$. The trace is $T=J_{33}-2f_*-\beta d_1^*<0$, and the determinant can be written as
\begin{equation}
\label{eq:det}
D=J_{33}(-1+\beta d_1^*f_*+f_*^2).
\end{equation}
Then, for point B to be stable the term inside parenthesis has to be positive. Although this is not a sufficient condition to prove that point B is stable, the numerical resolution of Eqs.~(\ref{c1})--(\ref{f}) shows that this is the case; this is the region where defectors are dominant. When the parenthesis in Eq.~(\ref{eq:det}) is negative point B becomes unstable, wich means that a small fraction of cooperators will grow and survive (notice that point B is the only fixed point with only defectors). Then, since point A is also unstable, in this situation there must exist a third (mixed) fixed point in the dynamics. Eq. (\ref{eq:mixed}) supplies the solution for this mixed fixed point and numerical solutions show it is a stable attractor, the one describing the stationary coexistence of cooperators and defectors found at large $\beta$ values. In order to obtain the cuve $\beta_c(q)$ separating the regions of dominance of defectors from the mixture of cooperators and defectors we should find $d_1^*$ from Eqs. (\ref{f}) and (\ref{d1bis}) and solve the equation 
\begin{equation}
\label{b2stable}
-1+\beta_c d_1^*f_*+f_*^2=0.
\end{equation}
The exact analytical solution of this transition curve is very cumbersome, so that we try two alternative routes. One is to obtain a numerical solution (see Fig.\ref{fig2}), the other one is to find an approximate analytical solution. In this sense, let us note that, if $\beta^2 q^3 \gg 1$ Eqs. (\ref{f}) and (\ref{d1bis}) show that $d_1^*\simeq (\beta q)^{-1}$, because $\beta q \gg f_* \simeq (\beta q^2)^{-1}$. In this limit, the instability condition (\ref{b2stable}) just gives
\begin{equation}
\label{beta}
\beta_c =q^{-3},
\end{equation}
which provides an excellent approximation not only for $\beta_c^2 q^3 \simeq q^{-3}\gg 1$ (say $q \lesssim 0.5$ ) but over the whole range $0\leq q\leq 1$ as shown when compared with the exact numerical solution (Fig.~\ref{fig2}).

Fig.~\ref{fig2} shows that cooperation is favored at large costs $q$ and large $\beta$, whereas defectors dominate in the opposite limit. The origin of the dependence on the average cost $q$ is rather direct: the larger the cost, the less favorable for defectors to reproduce. The dependence on parameter $\beta$ is, however, counterintuitive since (at first sight) one would expect that large attack rates (large $\beta$) should benefit defectors. The explanation is not easy due to the nonlinearities involved in the model. One might think that the origin of the observed behavior relies on the exploitation mechanism that explains the existence itself of coexistence states, and accordingly reason that large attacking rates would cause a great damage on cooperators, which would reduce rewards over costs, ultimately harming defectors. However, this is not what happens, since we have seen above that the average reward $E'_r$ is a function of $q$ only, and then it does not change when increasing $\beta$ at fixed $q$. 

One explanation of why large $\beta$ favor cooperators is that it leads to a small fraction of defectors in the active state ($D_1$), thus reducing the damage on cooperators. In effect, if resources are abundant individuals receive them frequently and there will be large populations of $D_1$ individuals; if resources are scarce, only a few individuals will be in state $E_i=1$. The same occurs if attacking rates are large. Since attacks are indiscriminate, defectors are also victim of the attacks, which decrease the number of $D_1$ individuals; conversely for small attacking rates. This explanation is consistent with the behavior of $d_1$ displayed in Fig.~\ref{fig1}a. Indeed $d_1$ decreases yet from $\beta=0$, i.e. below the transition, as it can be seen from our approximate solution $d_1^* \simeq (\beta q)^{-1}$. Below some critical population value depending on $q$ (around $q^{2}$) the reduced population of defectors in the active state is not capable of extinguishing cooperators. In this point is worthwhile to point out that parasites continuously receive resources from the environment and interact, and then, they change from active to inactive states continuosly. In the stationary state, the fraction of defectors in the population is $d_0+d_1$. These defectors spend a fraction of time $d_0/(d_0+d_1)$ in inactive states and $d_1/(d_0+d_1)$ in active states.

In summary, we have developed a simple model describing a phase transition from defective parasitic populations (which dissipate some amount of resources in order to gain a higher reproduction rate) when resources are abundant to the survival of cooperators when resources are scarce. This is the result of a  self-organizing process involving subtle nonlinearities in the interactions induced by resource constraints. In contrast to previous models, where the same limiting resource ruled reproduction and population size \cite{requejo:2012,requejo:2012b} and do not display this transition, the model studied here assumes that the factor limiting reproduction is different from the one limiting the population size, so that populazion size remains approximately constant. This may be accomplished in chemostat or retentostat experiments \cite{monod:1950,novik:1950,jannasch:1974,groisman:2005}, which may allow for laboratory testing of the predictions of the model. Indeed, recent experiments with \emph{S.Cerevisiae} \cite{jansen:2005} and \emph{E.Coli} \cite{notley-mcrobb:1999} at low concentration of glucose agree with the results presented here of the survival of cooperative traits instead of their expected extinction in unstructured populations.

In natural environments, the constant population assumption may apply in situations where space constraints the size of the population more restrictively than resource scarcity. Of course, a complete description of spatially distributed populations goes beyond the mean-field model presented here, and should consider that interactions occur only among neighbors. Note however, that space alone is well-known to favor cooperation because it permits the formation of clusters of cooperators. Since we have seen here that resource limitation alone already allows for the survival of cooperation, the combined effects of both space and resource limitations are expected to enhance the conditions under which cooperators can prevent their extinction. 
Other extensions of the model, beside the inclusion of space, may be the introduction of continuous behaviors (and not just two, namely cooperate or defect), what could shed light on the observed phenotypic radiation of behaviors in \emph{E.Coli} \cite{maharjan:2006}, which represents an exception to the competitive exclusion principle \cite{hardin:1960}. 

In a broader scope, the results presented here might have played a role in the route towards the emergence of multicellularity by cooperative aggregation triggered by resource constraints. It has been argued \cite{pfeiffer:2001,pfeiffer:2003,koschwanez:2011} that such transition happened whenever cooperative individuals formed clusters, which subsequently evolved nutrient exchange between the components of the cluster, and later evolved a joint replication mechanism, stage at which a higher-order organism can be considered to exist. However, such studies do not explain why cooperative bacteria survived in a first stage before forming clusters. The mechanism presented here provides some insights for the maintainance of such cooperative individals before clusters of cooperators could form, being a first step towards the formation of multicellular organisms.



This work has been supported by the Spanish government (FIS2009-13370-C02-01) and the Generalitat de Catalunya (2009SGR0164). R.J.R. acknowledges the financial support of the Universitat Autònoma de Barcelona and the Spanish government (FPU grant).


\end{document}